\documentclass[twocolumn]{openjournal}

\usepackage{lipsum}

\usepackage{xcolor}
\usepackage{textgreek}
\usepackage[utf8]{inputenc}
\usepackage[english]{babel}

\usepackage{tabularx}
\usepackage{array}
\newcolumntype{M}[1]{>{\centering\arraybackslash}m{#1}}
\usepackage{float}

\usepackage{hyperref}
\hypersetup{
    unicode, 
    colorlinks=true,
    linkcolor=linkcolor,
    citecolor=linkcolor,
    filecolor=linkcolor,
    urlcolor=linkcolor,
}
\usepackage{color,colortbl}
\definecolor{linkcolor}{rgb}{0.0,0.3,0.5}
\usepackage{tensind}
\tensordelimiter{?}
\DeclareGraphicsExtensions{.bmp,.png,.jpg,.pdf}
\usepackage{verbatim}
\usepackage[normalem]{ulem}
\usepackage{orcidlink}
\usepackage{soul}

\graphicspath{ {./figs/} }

\begin{document}
\title{Rock vapour is opaque: implications for dynamics and observations of lava planets}

\author{T. Giang Nguyen}
\affiliation{Department of Physics, McGill University, 3600 rue University, Montr\'eal, QC H3A 2T8, Canada}
\email{giang.nguyen@mail.mcgill.ca}  

\author{Nicolas B. Cowan}
\affiliation{Department of Physics, McGill University, 3600 rue University, Montr\'eal, QC H3A 2T8, Canada}
\affiliation{Department of Earth \& Planetary Sciences, 3450 rue University, Montr\'eal, QC H3A 0E8, Canada}

\author{Gunnar Montseny Gens}
\affiliation{Ann and H.J. Smead Department of Aerospace Engineering Sciences, University of Colorado Boulder, 3775 Discovery Drive, Boulder, CO 80303, USA}

\author{Charles-\'Edouard Boukar\'e}
\affiliation{Department of Physics and Astronomy, York University, 4700 Keele St, Toronto, Ontario, Canada}

\author{William Eaton}
\affiliation{Department of Physics, McGill University, 3600 rue University, Montr\'eal, QC H3A 2T8, Canada}

\author{Karolina Sienko}
\affiliation{Department of Physics, McGill University, 3600 rue University, Montr\'eal, QC H3A 2T8, Canada}

\begin{abstract}
Extreme instellation on lava planets causes the rocky surface to melt and vaporize. Because the rock vapour composition is intrinsically tied to the mantle, atmospheric characterization of lava planets can hold valuable insight into the interior processes of rocky planets. To help interpret current data and strategize for future observations, we develop the model \texttt{SonicVapour} to simulate the dynamics of chemically complex secondary atmosphere of lava planets. We find that for planets with surface temperatures exceeding 2700 K, the rock vapour outgassed is optically thick, making the atmosphere vertically isothermal thus suppressing convection and severely limiting atmospheric detection via emission spectroscopy. In contrast, cooler planets with surfaces between 2300 K -- 2700 K have an atmospheric opacity close to 50\% and produce distinct spectral features. Counter-intuitively, therefore, cooler lava planet atmospheres are easier to detect. Our results ultimately emphasize the importance of considering atmospheric ``detectability" in tandem with signal-to-noise for future observation programs.
\end{abstract}


\maketitle

\section{Introduction} \label{sec:intro}


Lava planets are rocky worlds with surface temperatures hot enough to melt their surface. Although planets such as Earth hosted global magma oceans shortly after formation, lava planets can sustain their magma oceans in perpetuity due to the sheer irradiation emitted by their host stars \citep{chao2021lava}. With such intense irradiation, lava planets are prone to lose their primordial atmospheres but a secondary atmosphere can appear through the vaporization of the magma surface \citep{lopez2012thermal}.

The secondary rock-vapour atmosphere of a lava planet is intricately tied to its interior, so atmospheric characterization may provide information about mantle composition and dynamics \citep{kite2016atmosphere,boukare2025solidification}. Lava planets therefore have been targets for many observation programs for both the Spitzer and James Webb Space Telescope: K2-141b \citep{zieba2022k2}, TOI-431b \citep{monaghan2025low}, 55-cnce  \citep{patel2024jwst}.


Atmospheric dynamics of a lava planet are unique and hard to simulate using existing frameworks. Lava planets are expected to be tidally-locked due to the close proximity to their host star, leading to a permanent dayside and nightside. This configuration can cause the secondary atmosphere to collapse on the nightside and the large pressure gradient between the two hemispheres can accelerate supersonic winds \citep{wordsworth2015atmospheric,nguyen2020modelling}.


Despite the challenges of developing an accurate model for lava planet atmospheres, there have been several advancements for collapsible atmospheres with self-consistent radiative transfer for lava planets \citep{castan2011atmospheres,nguyen2024clouds}. These works, however, are limited to an atmosphere consisting entirely of a single gas, e.g., Na or SiO. To fully realize the chemical complexity of lava planet atmospheres, we introduce an open-source code, \texttt{SonicVapor}\footnote{https://github.com/TueGiangNguyen/SonicVapour}, which calculates the steady-state atmospheric flow for any given magma ocean composition. We then use this model to simulate the dynamics and observations of hot and warm examples of lava planets: K2-141b and TOI-431b.

\section{Model description} \label{sec:model}

\subsection{Hydrodynamics}

\begin{figure}[ht]
    \centering
    \includegraphics[width=\linewidth]{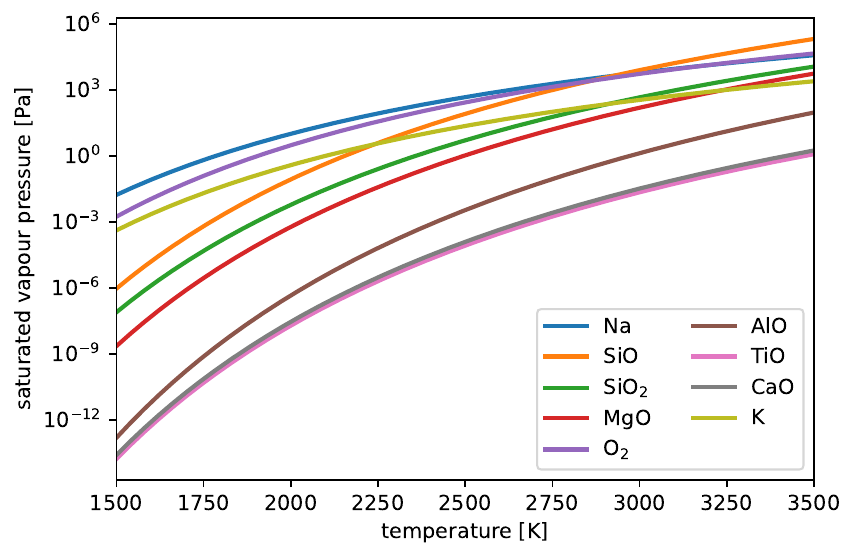}
    \caption{Saturated vapour pressure of nine most abundant species outgassed from Bulk Silicate Earth magma as predicted by \texttt{LavAtmos} \citep{van2023lavatmos}. The two most dominant species are Na which dominates at low temperatures, and SiO which dominates at high temperatures (\textgreater 2800 K).}
    \label{fig:pvap}
\end{figure}

As a first order approximation, we assume that the mantles of lava planets have Bulk Silicate Earth composition with 45\% SiO$_2$, 37\% MgO, 8\% FeO, and other trace elements \citep{holland2003treatise}. Given the magma composition, we then use the open-source code \texttt{LavAtmos} to find the saturated vapour pressure of the resulting outgassed atmosphere shown in Fig. \ref{fig:pvap} \citep{van2023lavatmos}.

For previous modelling efforts of atmospheres with a single constituent \citep{nguyen2022impact,nguyen2024clouds}, we isolated the saturated vapour pressure relation of either SiO or Na using chemical equilibrium models of \cite{schaefer2009chemistry} or \cite{van2023lavatmos}. To incorporate multiple outgassed species into the hydrodynamical framework, we must first define the total pressure, mean molecular mass, and mean specific heat capacity.

The total atmospheric pressure is simply the sum of the  partial pressures: $P_{\text{tot}} = \sum P_i$. The mean molecular mass is calculated via the partial pressures: $\bar{m} = (1/P_{\text{tot}})\sum P_i m_i$. The same is done for specific heat capacity: $\bar{C_p} = (1/P_{\text{tot}})\sum P_i (C_p)_i$. We can insert these terms into the generalized system of equations as defined by the turbulent boundary layer model of \cite{ingersoll1985supersonic} built to handle steady-state collapsible atmospheres:

\begin{equation}
    \partial_x \bigg(\frac{P_{\text{tot}}}{g} V\bigg) = \frac{P_{\text{tot,v}} - P_{\text{tot}}}{\sqrt{2 \pi R T_s}},
    \label{eq:massg}
\end{equation}
\begin{equation}
    \partial_x \bigg(\frac{P_{\text{tot}}}{g} V^2\bigg) = -\partial_x \int_z P_{\text{tot}} dz + \tau,
    \label{eq:momg}
\end{equation}
\begin{equation}
    \partial_x \bigg[\frac{P_{\text{tot}}}{g} V \left(\frac{V^2}{2} + \bar{C_p} T\right)\bigg] = Q,
    \label{eq:eneg}
\end{equation} \\

\noindent where $T$ and $T_s$ are atmospheric and surface temperature respectively, $P_{\text{tot,v}}$ is the sum of saturated partial pressures, $V$ is the wind speed, $g$ is gravity, $R = k/\bar{m}$ is the gas constant, $\tau$ is drag, and $Q$ is the energy balance of the atmosphere consisting of radiative flux and sensible heat flux. These equations are solved following \cite{nguyen2024clouds} and the reader can consult the appendix of that paper or the GitHub repository for how we calculate the fluxes and solve the system of equations. Solving for $P$, $T$, and $V$ yields the pressure, temperature, and wind speed at the boundary of the turbulent layer, which is at an altitude of half the scale height \citep{ingersoll1985supersonic}.

This new framework essentially treats an outgassed molecule as a singular species having the mean-weighted temperature-dependent properties of all the gas species. This assumes that the atmospheric mass fraction should follow the atmospheric abundances outputted by \texttt{LavAtmos}. Although \texttt{LavAtmos} outputs up to 30 gas species, we focus on the 9 species that are the most abundant and impactful for radiative transfer: Na, SiO, SiO$_2$, MgO, O$_2$, AlO, TiO, CaO, and K.

A drawback to this approach is that there is a net evaporative zone and condensation zone for the entire atmosphere whereas a realistic lava planet atmosphere should have different zones for each gas species (i.e SiO should condense at hotter temperatures than Na). Another limit is that Coriolis forcing is neglected due to the additional complexity of 3D flow instead of 2D (horizontal along co-latitudes and vertical). \cite{nguyen2024clouds} also included cloud formation and latent heat but these processes have only a minor impact on the atmosphere and are hard to implement for multi-component atmospheres; these processes are also neglected.

\subsection{Radiative transfer}


With the chemical complexity implemented into the hydrodynamical equations, we move onto the radiative transfer within the atmosphere. This is done by calculating heating by absorption of stellar and surface flux, and thermal emission of the atmosphere. The balance of these terms contribute towards solving $Q$ in Eq. \ref{eq:eneg} and is also performed following \cite{nguyen2024clouds}.

\begin{figure}[ht]
    \centering
    \includegraphics[width=\linewidth]{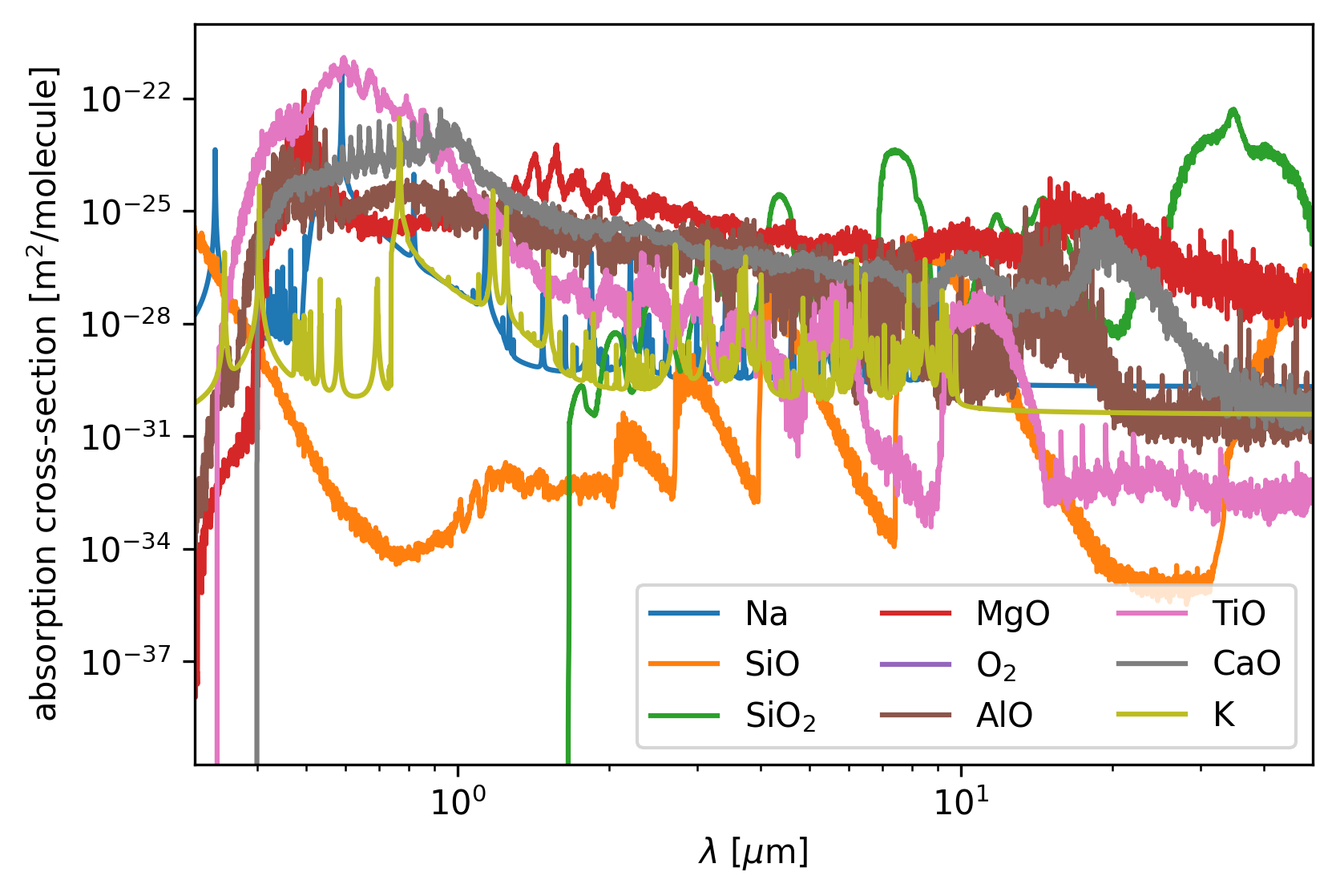}
    \caption{Absorption cross-section per molecule extracted from the ExoMol database. For this plot, we set the temperature to 2800 K and pressure to 7.7 kPa as thermal and pressure broadening parameters. Note that MgO has very broad peaks spanning the visible and IR and will play a large role in the atmosphere's radiative balance.}
    \label{fig:xcross}
\end{figure}

What is new for radiative transfer compared to previous work is that the atmosphere contains 9 gas species instead of being purely SiO. We extract absorption cross-section of each gas species from the database ExoMol \citep{tennyson2012exomol}, shown in Fig. \ref{fig:xcross}.

Solving Equations \ref{eq:massg} -- \ref{eq:eneg} lets us find the total pressure and atmospheric temperature. We then calculate the abundance of each species using the molar mass fraction outputs of \texttt{LavAtmos}. These terms are then used with the absorption cross-sections to compute the opacity spectrum for the appropriate pressure and temperature broadening. We take the sum of the optical depths calculated for each species to determine the spectral opacity, $\epsilon(\lambda)$, using \texttt{petitRADTRANS} \citep{molliere2019petitradtrans}.

Once the spectral opacity is found at a location, we calculate the radiative fluxes in the same manner as \cite{nguyen2024clouds}. Stellar absorption is computed as $\int \epsilon(\lambda) F_*(\lambda) d\lambda$ where $F_*$ is the stellar spectrum. Surface flux absorption is computed as $\pi\int \epsilon(\lambda) B(\lambda,T_s) d\lambda$ while radiative cooling is  $\pi\int\epsilon(\lambda) B(\lambda,T) d\lambda$ where $B$ is the Planck function.

Stellar data are derived from analogous stars from the MUSCLES survey \citep{france2016muscles}: we scale the spectrum of HD85512 for K2-141 and HD40307 for TOI-431. Because of the high opacity in the visible, we also impose a vertically isothermal profile for the purposes of solving the fluid dynamics \citep{chandrasekhar1947transfer,chandrasekhar1960radiative}. This is also consistent with other radiative transfer simulations of lava planets \citep{zilinskas2022observability}.

\section{Results} \label{sec:results}

\subsection{Model results}

\begin{figure}[ht]
    \centering
    \includegraphics[width=\linewidth]{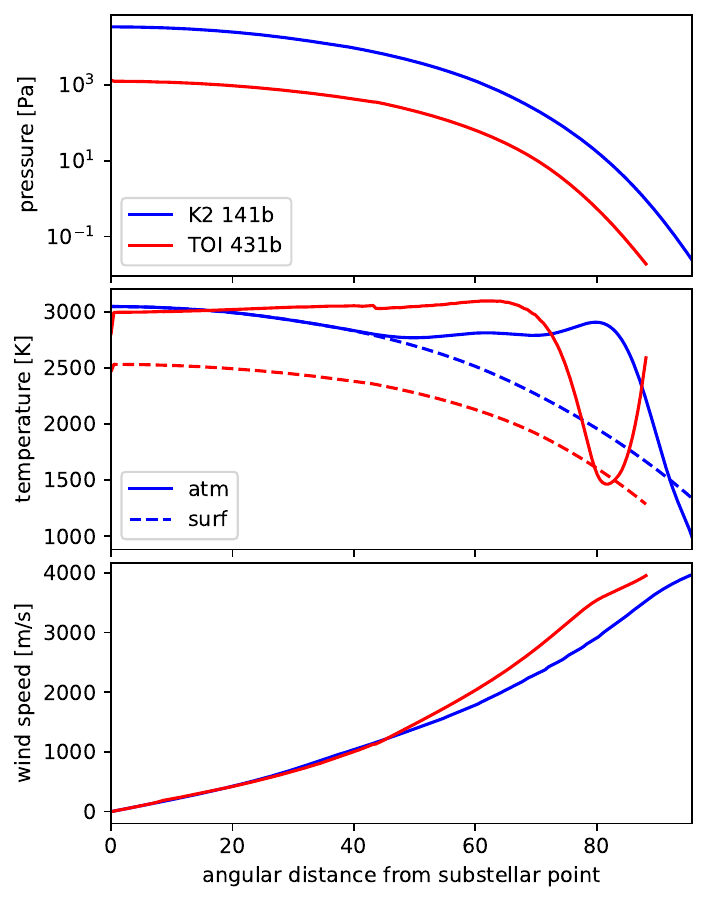}
    \caption{Atmospheric pressure (\textit{top}), surface and atmospheric temperature (\textit{middle}), and wind speeds (\textit{bottom}) of K2-141b and TOI-431b at an altitude of half a scale-height. Because the surface temperature of K2-141b is hundreds of Kelvin hotter, the atmosphere of K2-141b is an order of magnitude thicker than that of TOI-431b. For K2-141b, the atmosphere shares the same temperature as the surface for much of the dayside due to the atmosphere's optical thickness. TOI-431b has an optically thinner atmosphere so there is an appreciable difference in temperature between the atmosphere and the surface. Surprisingly, both planets have similar wind speeds despite having wildly different atmospheric conditions.}
    \label{fig:PTV}
\end{figure}

We used the methods described above to model the atmospheric conditions of K2-141b and TOI-431b. K2-141b is our the high-temperature case with a massive and optically thick atmosphere. TOI-431b is the low-temperature case with a much thinner atmosphere. Basic planetary parameters are shown in Table \ref{tab:planet}. The model results are shown in Figure. \ref{fig:PTV}.

\begin{table}[ht]
\centering
\begin{tabular}{p{0.25\columnwidth}|M{0.30\columnwidth}|M{0.30\columnwidth}}

        & K2-141b & TOI-431b  \\ \hline
      Planet radius  & 1.54 R$_\oplus$ & 1.28 R$_\oplus$ \\ \hline
      Stellar radius & 0.681 R$_*$ & 0.731 R$_*$ \\ \hline
      Distance & 2.30 a/R$_*$ & 3.32 a/R$_*$ \\ \hline
      Stellar temp. & 4373 K & 4850 K \\ \hline
      Mass & 5.08 M$_E$ & 3.07 M$_E$ \\ \hline
      Irr. temp. & 3056 K & 2466 K \\ \hline
      Source & \cite{malavolta2018ultra} & \citep{osborn2021toi} \\
\end{tabular}
\caption{Basic properties of lava planets K2-141$\rm{b}$ and TOI-431$\rm{b}$. Despite orbiting a cooler star, K2-141$\rm{b}$'s proximity to its host makes it hotter.}
\label{tab:planet}
\end{table}


For K2-141b, the opacity is near unity within 40$^\circ$ of the sub-stellar point so the atmosphere essentially radiates like a perfect blackbody. The atmospheric temperature is the same as the surface temperature --- vertically isothermal --- because the atmosphere radiates to the surface as much as the surface radiates to the atmosphere. Beyond 40$^\circ$, the atmosphere becomes optically thin enough that atmospheric temperature rises above surface temperatures for reasons described in Section \ref{sec:imp}.


For TOI-431b, its atmosphere is an order of magnitude less massive than K2-141b's. Because TOI-431b's atmosphere never approaches an opacity of unity, the surface is significantly cooler than the atmosphere. When comparing to previous results of TOI-431b from \cite{nguyen2024clouds} with an optically thin pure SiO atmosphere, the thicker atmosphere in this work cools the surface by $\sim$200 K because less visible light hits the surface.

\subsection{Synthetic eclipse spectra}

\begin{figure}
    \centering
    \includegraphics[width=\linewidth]{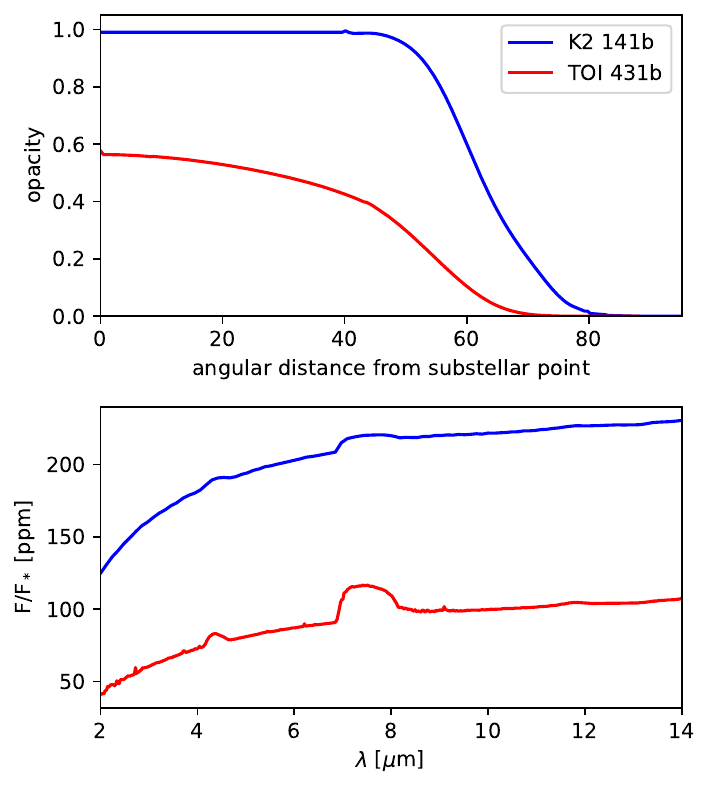}
    \caption{Total opacity (\textit{top}) and synthetic eclipse spectra of K2-141b and TOI-431b (\textit{bottom}). Total opacity is calculated as the fraction of stellar absorption and total bolometric flux. Because most of K2-141b's opacity is roughly unity, the atmosphere radiates like a perfect blackbody resulting in severely muted spectral features. TOI-431b, on the other hand, has significantly lower opacity for its entire atmosphere making atmospheric signals more distinguishable in its eclipse spectrum.}
    \label{fig:eclipses}
\end{figure}

Given the spatial distribution of surface and atmospheric temperatures, we construct a global map of outgoing top-of-atmosphere emission to create a synthetic eclipse emission spectrum. This process is also identical to \cite{nguyen2024clouds} which itself is based on \cite{cowan2008inverting}. The eclipse spectra are shown in Fig. \ref{fig:eclipses}.

For K2-141b, as seen in Fig. \ref{fig:eclipses}, the opacity is near unity until an angular distance of 40$^\circ$. In this region, the atmosphere and surface have the same temperature and so result in a blackbody spectrum devoid of emission or absorption features. Because distinguishable atmospheric emission comes from only about half of the dayside hemisphere, K2-141b's eclipse spectrum shows very muted atmospheric spectral features.


For TOI-431b, the opacity never exceeds 0.6 so atmospheric temperature and surface temperature differ significantly with the atmosphere being much hotter than the surface. Unlike K2-141b, the entirety of TOI-431b's dayside hemisphere produces an emission feature so its eclipse spectrum shows much more pronounced atmospheric spectral features.

\section{Implications} \label{sec:imp}

\subsection{Dynamics}

\begin{figure}[ht]
    \centering
    \includegraphics[width=\linewidth]{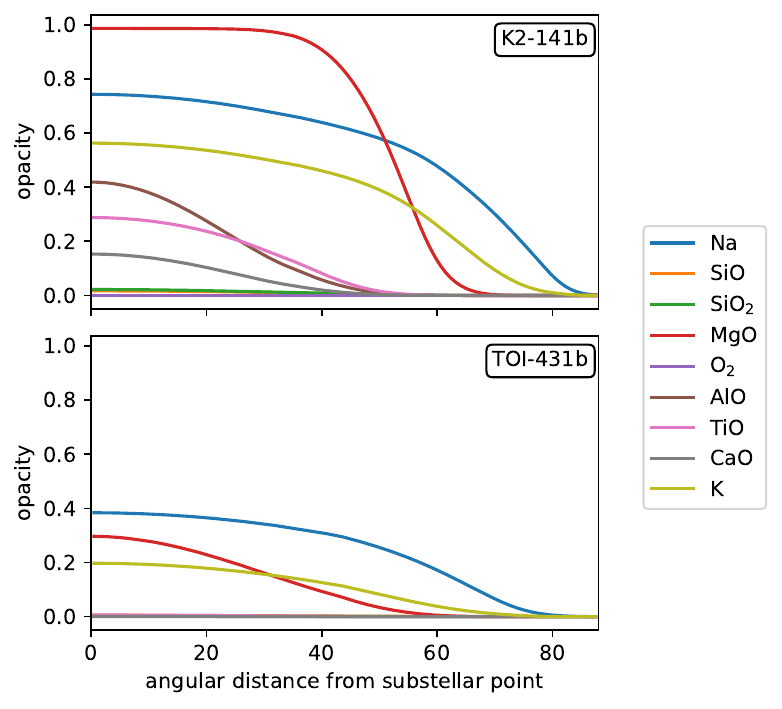}
    \caption{Opacity contribution of individual gas species on K2-141b (top) and TOI-431b (bottom). For thin atmospheres like TOI-431b with substellar surface pressures of $\sim10^3$ Pa, Na and K are the main contributor of stellar flux absorption. Given a thick enough atmosphere like K2-141b's, there is enough MgO ($\sim 10^2$ Pa) to absorb all incoming stellar radiation.}
    \label{fig:spec_opac}
\end{figure}

To gauge the radiative impact of atmospheric composition and abundances, we plot the opacity contributions of each gas species in Fig. \ref{fig:spec_opac}. Sodium is the most volatile gas species studied and its strong spectral features in the visible contribute to significant atmospheric heating since most of the starlight is visible light. Potassium also has strong spectral features in the visible but is not as abundant as Na. Therefore, the contribution to opacity by K follows the same trend as Na but is generally 30\% lower.

An unexpected major contributor to opacity is MgO. It has relatively weak spectral features but the features are broad, encompassing much of the visible and IR. For TOI-431b, MgO abundances are not high enough for lines to be saturated. Conversely, K2-141b has enough atmospheric MgO to saturate IR and visible lines, which makes the atmosphere completely opaque at all wavelengths. Because other gases have relatively narrow spectral peaks, MgO dominates atmospheric opacity in high-temperature lava planets.

For hot planets with substellar temperatures around 3000 K such as K2-141b, their atmospheres are virtually opaque. Dynamically, this results in a vertically isothermal atmosphere around the substellar region which transitions to a temperature inversion towards the terminator. For cooler planets with surface substellar temperatures around 2500 K like TOI-431b, the atmosphere should have an inversion everywhere on the planet, consistent with \cite{zilinskas2022observability}.

\cite{nguyen2022impact} showed that when the atmosphere has relatively inefficient IR cooling, temperatures must rise to emit more shortwave radiation and maintain radiative balance. Because the optical depth of rock vapour is strongest in the visible, optically thin atmospheres become hotter than the surface thus producing a significant temperature inversion. This suggests that convection is severely inhibited for cool lava planets implying that these atmospheres stratify easily and cloud formation is suppressed. In contrast, optically thick atmospheres from hotter lava planets are isothermal and thus more convectively stable, leading to more opportunities for heterogeneous cloud nucleation.

\subsection{Observability}

\begin{figure}
    \centering
    \includegraphics[width=\linewidth]{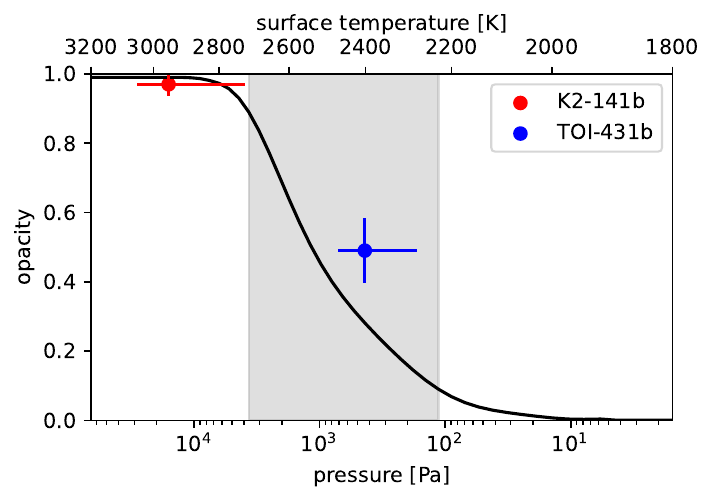}
    \caption{Opacity as a function of surface temperature and pressure determined via the stellar spectrum of K2-141. The grey area highlights where opacity is between 0.1 and 0.9. The red and blue lines show the opacity and pressure range for the substellar region ($\theta <$ 50$^\circ$) of K2-141b and TOI-431b. For opacity greater than 0.9, the atmosphere absorbs and reradiates nearly all fluxes which makes the atmosphere isothermal and inhibits atmospheric detection. For opacity lower than 0.1, the atmosphere is nearly transparent at all wavelengths, which also inhibits spectral features. Therefore, planets with surface temperatures in the grey area (2300 -- 2700 K) are optimal for atmospheric detection and characterization.}
    \label{fig:gen_opacity}
\end{figure}

The conditions around the substellar region that dictate atmospheric dynamics and the opacity in this region is the most important factor for atmospheric detection. An optically thin atmosphere will allow all light to pass through it unimpeded so thermal emission will come from the surface, which should behave like a perfect blackbody given the low albedo of magma \citep{essack2020low}. Conversely, atmospheric opacity near unity will also yield blackbody-like emission. Therefore, the atmospheric characterization window lies at opacities of 0.1 -- 0.9 corresponding to surface temperatures between 2300 -- 2700 K as seen in Fig. \ref{fig:gen_opacity}. 

Our results highlight a paradox for atmospheric characterization of lava planets: it is harder to detect a thick atmosphere than a thin atmosphere. As seen in the eclipse spectra of Fig. \ref{fig:eclipses}, the hotter K2-141b with 10 times the atmospheric mass shows weaker atmospheric signals than the cooler TOI-431b. This has profound implications for observation strategies.

K2-141b is a popular target for observations because of its high signal-to-noise ratio \citep{dang2021hell,espinoza2021first}. It is also very hot which reinforces confidence that there is an appreciable atmosphere present to study. However, our work shows that target selection should consider the optical thickness of the atmosphere when the goal is to detect the spectral signature of an atmosphere.

\section{Conclusions} \label{sec:conclusions}


We improved upon past work to build \texttt{SonicVapour}, an open-source code that models the chemically-complex steady-state hydrodynamics of lava planet atmospheres. We then simulate the atmospheres of K2-141b and TOI-431b to test the limiting cases of optically thick and thin atmospheres.

Our results showed that for a thin atmosphere ($\sim 10^3$ Pa), Na and K dominate stellar absorption and the atmosphere exhibits emission features. With a thicker atmosphere ($\sim 10^4$ Pa) like K2-141b, the trace gas MgO becomes abundant enough to dominate radiative transfer and the atmosphere becomes opaque at all wavelengths.


For an opaque atmosphere, top-of-atmosphere emission essentially looks like a perfect blackbody which is indistinguishable from the magma. Counter-intuitively, an atmosphere needs to be thin enough to produce a significant atmospheric signal. This result suggests a paradigm shift for lava planet observations which should focus on somewhat cooler planets to characterize their atmospheres.


There are many avenues to further use and expand \texttt{SonicVapour}. Although we used a Bulk Silicate Earth composition, this could be expanded to any magma composition admitted by chemical equilibrium models like \texttt{LavAtmos}. Therefore, one could use interior models from \cite{boukare2025solidification} or \cite{nicholls2024magma} to determine the changes in atmospheric conditions as the magma ocean cools and differentiates. One could also arbitrarily remove volatiles like Na from the atmosphere to mimic cold-trapping \citep{nguyen2020modelling} or hydrodynamical escape \citep{ito2021hydrodynamic}. One could also implement Coriolis forces given the results of \cite{lai2024ocean} to better spatially resolve the atmosphere. These improvements would improve the fidelity of our simulations of lava planet atmospheres.

\section*{Acknowledgments}

This research was partially funded by the Canadian Space Agency's James Webb Space Telescope observer's program and the Heising-Simons Foundation. N.B.C. acknowledges support from an NSERC Discovery Grant, a Tier 2 Canada Research Chair, and an Arthur B. McDonald Fellowship. The authors also thank the Trottier Space Institute and l’Institut de recherche sur les exoplan\`etes for their financial support and dynamic intellectual environment.

\bibliographystyle{aasjournal}
\bibliography{oja_template}

@misc{van2023lavatmos,
  title={LavAtmos: An open-source chemical equilibrium vaporization code for lava worlds},
  author={Van Buchem, Christiaan PA and Miguel, Yamila and Zilinskas, Mantas and van Westrenen, Wim},
  year={2023},
  publisher={Wiley Online Library}
}

@article{chao2021lava,
  title={Lava worlds: From early earth to exoplanets},
  author={Chao, Keng-Hsien and deGraffenried, Rebecca and Lach, Mackenzie and Nelson, William and Truax, Kelly and Gaidos, Eric},
  journal={Geochemistry},
  volume={81},
  number={2},
  pages={125735},
  year={2021},
  publisher={Elsevier}
}

@article{lopez2012thermal,
  title={How thermal evolution and mass-loss sculpt populations of super-Earths and sub-Neptunes: application to the Kepler-11 system and beyond},
  author={Lopez, Eric D and Fortney, Jonathan J and Miller, Neil},
  journal={The Astrophysical Journal},
  volume={761},
  number={1},
  pages={59},
  year={2012},
  publisher={IOP Publishing}
}

@article{kite2016atmosphere,
  title={Atmosphere-interior exchange on hot, rocky exoplanets},
  author={Kite, Edwin S and Fegley Jr, Bruce and Schaefer, Laura and Gaidos, Eric},
  journal={The Astrophysical Journal},
  volume={828},
  number={2},
  pages={80},
  year={2016},
  publisher={IOP Publishing}
}

@article{zieba2022k2,
  title={K2 and Spitzer phase curves of the rocky ultra-short-period planet K2-141 b hint at a tenuous rock vapor atmosphere},
  author={Zieba, Sebastian and Zilinskas, Mantas and Kreidberg, Laura and Nguyen, Tue Giang and Miguel, Yamila and Cowan, Nicolas B and Pierrehumbert, Ray and Carone, Ludmila and Dang, Lisa and Hammond, Mark and others},
  journal={Astronomy \& Astrophysics},
  volume={664},
  pages={A79},
  year={2022},
  publisher={EDP Sciences}
}

@article{monaghan2025low,
  title={Low 4.5 $\mu$m Dayside Emission Disfavors a Dark Bare-rock Scenario for the Hot Super-Earth TOI-431 b},
  author={Monaghan, Christopher and Roy, Pierre-Alexis and Benneke, Bj{\"o}rn and Crossfield, Ian JM and Coulombe, Louis-Philippe and Piaulet-Ghorayeb, Caroline and Kreidberg, Laura and Dressing, Courtney D and Kane, Stephen R and Dragomir, Diana and others},
  journal={The Astronomical Journal},
  volume={169},
  number={5},
  pages={239},
  year={2025},
  publisher={IOP Publishing}
}

@article{patel2024jwst,
  title={JWST reveals the rapid and strong day-side variability of 55 Cancri e},
  author={Patel, Jayshil Ashokkumar and Brandeker, Alexis and Kitzmann, D and dit de la Roche, DJM Petit and Bello-Arufe, A and Heng, K and Vald{\'e}s, E Meier and Persson, CM and Zhang, M and Demory, B-O and others},
  journal={Astronomy \& Astrophysics},
  volume={690},
  pages={A159},
  year={2024},
  publisher={EDP Sciences}
}

@article{wordsworth2015atmospheric,
  title={Atmospheric heat redistribution and collapse on tidally locked rocky planets},
  author={Wordsworth, Robin},
  journal={The Astrophysical Journal},
  volume={806},
  number={2},
  pages={180},
  year={2015},
  publisher={IOP Publishing}
}

@article{nguyen2020modelling,
  title={Modelling the atmosphere of lava planet K2-141b: implications for low-and high-resolution spectroscopy},
  author={Nguyen, T Giang and Cowan, Nicolas B and Banerjee, Agnibha and Moores, John E},
  journal={Monthly Notices of the Royal Astronomical Society},
  volume={499},
  number={4},
  pages={4605--4612},
  year={2020},
  publisher={Oxford University Press}
}

@article{nguyen2022impact,
  title={The impact of ultraviolet heating and cooling on the dynamics and observability of lava planet atmospheres},
  author={Nguyen, T Giang and Cowan, Nicolas B and Pierrehumbert, Raymond T and Lupu, Roxana E and Moores, John E},
  journal={Monthly Notices of the Royal Astronomical Society},
  volume={513},
  number={4},
  pages={6125--6133},
  year={2022},
  publisher={Oxford University Press}
}

@article{nguyen2024clouds,
  title={Clouds in Partial Atmospheres of Lava Planets and Where to Find Them},
  author={Nguyen, T Giang and Cowan, Nicolas B and Dang, Lisa},
  journal={The Astronomical Journal},
  volume={168},
  number={6},
  pages={287},
  year={2024},
  publisher={IOP Publishing}
}

@book{holland2003treatise,
  title={Treatise on geochemistry: 8. Biogeochemistry},
  author={Holland, Heinrich D and Turekian, Karl K},
  year={2003}
}

@article{schaefer2009chemistry,
  title={Chemistry of silicate atmospheres of evaporating super-Earths},
  author={Schaefer, Laura and Fegley, Bruce},
  journal={The Astrophysical Journal},
  volume={703},
  number={2},
  pages={L113},
  year={2009},
  publisher={IOP Publishing}
}

@article{ingersoll1985supersonic,
  title={Supersonic meteorology of Io: Sublimation-driven flow of SO2},
  author={Ingersoll, Andrew P and Summers, Michael E and Schlipf, Steve G},
  journal={Icarus},
  volume={64},
  number={3},
  pages={375--390},
  year={1985},
  publisher={Elsevier}
}

@article{lai2024ocean,
  title={Ocean circulation on tide-locked lava worlds. ii. scalings},
  author={Lai, Yanhong and Kang, Wanying and Yang, Jun},
  journal={The Planetary Science Journal},
  volume={5},
  number={9},
  pages={205},
  year={2024},
  publisher={IOP Publishing}
}

@article{tennyson2012exomol,
  title={ExoMol: molecular line lists for exoplanet and other atmospheres},
  author={Tennyson, Jonathan and Yurchenko, Sergei N},
  journal={Monthly Notices of the Royal Astronomical Society},
  volume={425},
  number={1},
  pages={21--33},
  year={2012},
  publisher={Blackwell Science Ltd Oxford, UK}
}

@article{molliere2019petitradtrans,
  title={petitRADTRANS-A Python radiative transfer package for exoplanet characterization and retrieval},
  author={Molli{\`e}re, P and Wardenier, JP and Van Boekel, R and Henning, Th and Molaverdikhani, K and Snellen, IAG},
  journal={Astronomy \& Astrophysics},
  volume={627},
  pages={A67},
  year={2019},
  publisher={EDP Sciences}
}

@article{france2016muscles,
  title={The MUSCLES treasury survey. I. Motivation and overview},
  author={France, Kevin and Loyd, RO Parke and Youngblood, Allison and Brown, Alexander and Schneider, P Christian and Hawley, Suzanne L and Froning, Cynthia S and Linsky, Jeffrey L and Roberge, Aki and Buccino, Andrea P and others},
  journal={The Astrophysical Journal},
  volume={820},
  number={2},
  pages={89},
  year={2016},
  publisher={IOP Publishing}
}

@article{zilinskas2022observability,
  title={Observability of evaporating lava worlds},
  author={Zilinskas, Mantas and Van Buchem, CPA and Miguel, Yamila and Louca, Amy and Lupu, Roxana and Zieba, Sebastian and van Westrenen, Wim},
  journal={Astronomy \& Astrophysics},
  volume={661},
  pages={A126},
  year={2022},
  publisher={EDP Sciences}
}

@article{cowan2008inverting,
  title={Inverting phase functions to map exoplanets},
  author={Cowan, Nicolas B and Agol, Eric},
  journal={The Astrophysical Journal},
  volume={678},
  number={2},
  pages={L129},
  year={2008},
  publisher={IOP Publishing}
}

@article{essack2020low,
  title={Low-albedo surfaces of lava worlds},
  author={Essack, Zahra and Seager, Sara and Pajusalu, Mihkel},
  journal={The Astrophysical Journal},
  volume={898},
  number={2},
  pages={160},
  year={2020},
  publisher={IOP Publishing}
}

@article{dang2021hell,
  title={A hell of a phase curve: Mapping the surface and atmosphere of a lava planet k2-141b},
  author={Dang, Lisa and Cowan, Nicolas B and Hammond, Mark and Kreidberg, Laura and Lupu, Roxana and Miguel, Yamila and Nguyen, Giang and Pierrehumbert, Raymond and Zieba, Sebastian and Zilinskas, Mantas},
  journal={JWST Proposal. Cycle 1},
  pages={2347},
  year={2021}
}

@article{espinoza2021first,
  title={The first near-infrared spectroscopic phase-curve of a super-Earth},
  author={Espinoza, Nestor and Bello-Arufe, Aaron and Buchhave, Lars A and Burgasser, Adam J and Demory, Brice-Olivier and Diamond-Lowe, Hannah and Fisher, Chloe and Gibson, Neale and Guzman Mesa, Andrea and Heng, Kevin and others},
  journal={JWST Proposal. Cycle 1},
  pages={2159},
  year={2021}
}

@article{boukare2025solidification,
  title={Solidification of Earth’s mantle led inevitably to a basal magma ocean},
  author={Boukar{\'e}, Charles-{\'E}douard and Badro, James and Samuel, Henri},
  journal={Nature},
  pages={1--6},
  year={2025},
  publisher={Nature Publishing Group UK London}
}

@article{nicholls2024magma,
  title={Magma ocean evolution at arbitrary redox state},
  author={Nicholls, Harrison and Lichtenberg, Tim and Bower, Dan J and Pierrehumbert, Raymond},
  journal={Journal of Geophysical Research: Planets},
  volume={129},
  number={12},
  pages={e2024JE008576},
  year={2024},
  publisher={Wiley Online Library}
}

@article{ito2021hydrodynamic,
  title={Hydrodynamic escape of mineral atmosphere from hot rocky exoplanet. I. Model description},
  author={Ito, Yuichi and Ikoma, Masahiro},
  journal={Monthly Notices of the Royal Astronomical Society},
  volume={502},
  number={1},
  pages={750--771},
  year={2021},
  publisher={Oxford University Press}
}

@article{malavolta2018ultra,
  title={An ultra-short period rocky super-Earth with a secondary eclipse and a Neptune-like companion around K2-141},
  author={Malavolta, Luca and Mayo, Andrew W and Louden, Tom and Rajpaul, Vinesh M and Bonomo, Aldo S and Buchhave, Lars A and Kreidberg, Laura and Kristiansen, Martti H and Lopez-Morales, Mercedes and Mortier, Annelies and others},
  journal={The Astronomical Journal},
  volume={155},
  number={3},
  pages={107},
  year={2018},
  publisher={IOP Publishing}
}

@article{osborn2021toi,
  title={TOI-431/HIP 26013: a super-Earth and a sub-Neptune transiting a bright, early K dwarf, with a third RV planet},
  author={Osborn, Ares and Armstrong, David J and Cale, Bryson and Brahm, Rafael and Wittenmyer, Robert A and Dai, Fei and Crossfield, Ian JM and Bryant, Edward M and Adibekyan, Vardan and Cloutier, Ryan and others},
  journal={Monthly Notices of the Royal Astronomical Society},
  volume={507},
  number={2},
  pages={2782--2803},
  year={2021},
  publisher={Oxford University Press}
}

@article{castan2011atmospheres,
  title={Atmospheres of hot super-Earths},
  author={Castan, Thibaut and Menou, Kristen},
  journal={The Astrophysical Journal Letters},
  volume={743},
  number={2},
  pages={L36},
  year={2011},
  publisher={IOP Publishing}
}

@article{chandrasekhar1947transfer,
  title={The transfer of radiation in stellar atmospheres},
  author={Chandrasekhar, Subrahmanyan},
  journal={Bulletin of the American Mathematical Society},
  volume={53},
  number={7},
  pages={641--711},
  year={1947}
}

@book{chandrasekhar1960radiative,
  title={Radiative transfer},
  author={Chandrasekhar, Subrahmanyan},
  year={1960},
  publisher={Courier Corporation}
}

\end{document}